\documentclass[mathleft
]{an_a_astroph}
\usepackage{graphicx}
\usepackage{times}
\usepackage{natbib}

\newcommand{\BV}{Brunt-V\"ais\"al\"a }

\newcommand{\msol}{\mbox{${\rm M}_{\odot}\;$}}

\newcommand{\msun}{\mbox{${\rm M}_{\odot}$}}

\begin{document}

\Pagespan{1}{}
\Yearpublication{2006}%
\Yearsubmission{2005}%
\Month{11}%
\Volume{999}%
\Issue{88}%

\title{Gravity modes and mixed modes as probes of stellar cores in main-sequence stars: from solar-like to $\beta$ Cep stars}

\author{A. Miglio \and J. Montalb\'an \and P. Eggenberger \and A. Noels}
\titlerunning{Gravity modes and mixed modes as probes of stellar cores in main-sequence stars}
\authorrunning{A. Miglio et al.}
\institute{
Institut d'Astrophysique et de G\'eophysique de l'Universit\'e de Li\`ege,
All\'ee du 6 Ao\^ut, 17 B-4000 Li\`ege, Belgium}



\abstract{%
We investigate how the frequencies of gravity modes depend on the detailed properties of the chemical composition gradient that develops near the core of main-sequence stars and, therefore, on the transport processes that are able to modify the $\mu$ profile in the central regions.
We show that in main-sequence models, similarly to the case of white dwarfs, the periods of high-order gravity modes are accurately described by a uniform period spacing superposed to an oscillatory component. The periodicity and amplitude of such component are related, respectively, to the location and sharpness of the $\mu$ gradient. We briefly discuss and interpret, by means of this simple approximation, the effect of turbulent mixing near the core on the periods of both high-order and low-order g modes, as well as of modes of mixed pressure-gravity character.
}

\maketitle

\section{Introduction}
In main sequence stars the periods of gravity modes are sensitive probes of stellar cores and, in particular, of the chemical composition gradient that develops near the outer edge of the convective core.

High-order gravity modes are observed in two classes of main-sequence stars: $\gamma$ Doradus ($M\simeq1.4-1.8$ \msun, see e.g. \citealt{Handler99} for a review) and Slowly Pulsating B stars, with masses from about 3 to 8 \msol and spectral type B3-B8 \citep{Waelkens91}.
The seismic modelling of these classes of pulsators is a formidable task to undertake. The frequencies of high-order g modes are in fact closely spaced  and can be severely perturbed by the effects of rotation \citep[see e.g.][]{Dintrans00,Suarez05}. Nonetheless, the high scientific interest of these classes of pulsators has driven efforts in both the observational and theoretical domains.
As suggested by \citet{Suarez05} in the case of $\gamma$ Dor stars, a seismic analysis becomes feasible for slowly rotating targets. In these favorable cases  the first-order asymptotic approximation \citep{Tassoul80} can be used as a tool to derive the buoyancy radius of the star \citep[see][]{Moya05} from the observed frequencies.
The g-mode spectra of these stars contain however much more information on the internal structure of the star: this is illustrated in Sec. \ref{sec:high} by means of a simple refinement of the asymptotic approximation of \citet{Tassoul80}.

In Sec. \ref{sec:low} we briefly discuss the case of low-order g modes, and modes of mixed pressure and gravity character.
In main-sequence stars these modes are detected in $\delta$ Scuti ($M\simeq1.5-2.5$ \msun, see e.g. the overview by \citealt{Rodriguez01}) and $\beta$ Cephei ($M\simeq7-20$ \msun, \citealt{Stankov05}) pulsators. Moreover solar-like oscillations in subgiant stars can also present a p-mode spectrum perturbed by the interaction between p and g modes.
In Sec. \ref{sec:low} we give an example on how different mixing processes acting near the core affect differently the oscillation spectra observed in these stars.
\section{High-order g modes}
\label{sec:high}
As it is well known, the period spectrum of gravity modes is determined by the spatial distribution of the \BV frequency ($N$) which is defined as:

\begin{equation}
N^2=g\left(\frac{1}{\Gamma_1 p}\frac{{\rm d}p}{{\rm d}r}-\frac{1}{\rho}\frac{{\rm d}\rho}{{\rm d}r}\right)\,{\rm ,}
\end{equation}
where $\rho$ is density, $p$ is pressure, $g$ is the gravitational acceleration and $\Gamma_1=\left(\partial\ln{p}/\partial\ln{\rho}\right)_{\rm ad}$.

$N$ can be approximated, assuming the ideal gas law for a fully-ionized gas, as:

\begin{equation}
N^2\simeq \frac{g^2\rho}{p}\left(\nabla_{\rm ad}-\nabla+\nabla_\mu \right)\,{\rm ,}
\label{eq:bvgr}
\end{equation}

\noindent
where

$$
\nabla=\frac{{\rm d}\ln{T}}{{\rm d} \ln{p}},\;\;\nabla_{\rm ad}=\left(\frac{\partial\ln{T}}{\partial\ln{p}}\right)_{\rm ad}\;\;{\rm and}\;\; \nabla_\mu=\frac{{\rm d}\ln{\mu}}{{\rm d}\ln{p}} \;{\rm ,}
$$
$T$ is temperature and $\mu$ is the mean molecular weight.

The term $\nabla_\mu$ gives the explicit contribution of a change of chemical composition to $N$.
The first order asymptotic approximation developed by \citet{Tassoul80} shows that, in the case of a model that consists of an inner convective core and an outer radiative envelope  \citep[we refer to the work by][ for a complete analysis of other possible cases]{Tassoul80}, the periods of low-degree, high-order g modes (where $N^2 \gg \omega^2$) are given by:

\begin{equation}
P_k=\frac{\pi^2}{L\,\int_{x_0}^{1}{\frac{|N|}{x} {\rm d}x}} \left(2k+n_e\right)\;{\rm ,}
\label{eq:asy}
\end{equation}

\noindent
where $L=[\ell(\ell+1)]^{1/2}$ (with $\ell$ the mode degree), $n_e$ the effective polytropic index of the surface layer, $x$ the normalized radius and $x_0$ corresponds to the boundary of the convective core. In order to avoid confusion with $n_e$, the radial order of g modes is represented by $k$.

Following Eq. \ref{eq:asy}, the periods are asymptotically equally spaced in $k$ and the spacing decreases with increasing $L$. It is therefore natural to introduce, in analogy to the large frequency separation of p modes, the {\it period spacing} of gravity modes, defined as:
\begin{equation}
\Delta P=P_{k+1}-P_{k}\;{\rm .}
\end{equation}
In this work we show that deviations from a constant $\Delta P$ contain information on the chemical composition gradient left by a convective core evolving on the main sequence.

The signatures of chemical stratifications in high-order gravity modes have been extensively investigated theoretically, and then observed, in pulsating white dwarfs \citep[see e.g.][for a review]{Kawaler95}.
The influence of chemical composition gradients on g modes in main sequence stars has been partly addressed and suggested in two farsighted works by \citet{Berthomieu88} and \citet{Dziembowski93}.
Inspired by these works and following the approach of \citet{Berthomieu88} and \citet{Brassard92} we investigate the properties of high-order, low-degree gravity modes in  main sequence stellar models.

\begin{figure}
\begin{center}
\resizebox{0.48\hsize}{!}{\includegraphics[angle=0]{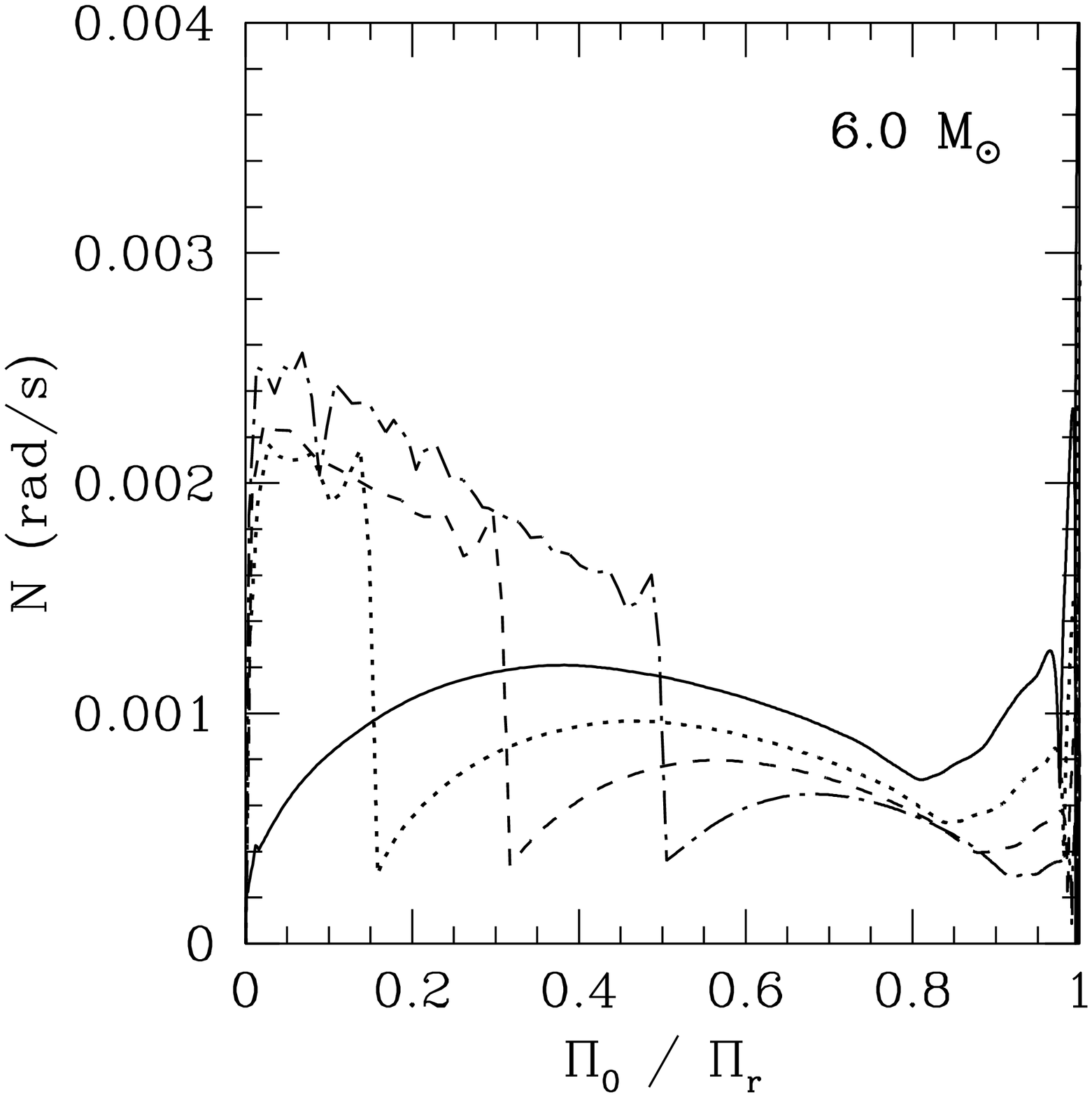}}
\resizebox{0.48\hsize}{!}{\includegraphics[angle=0]{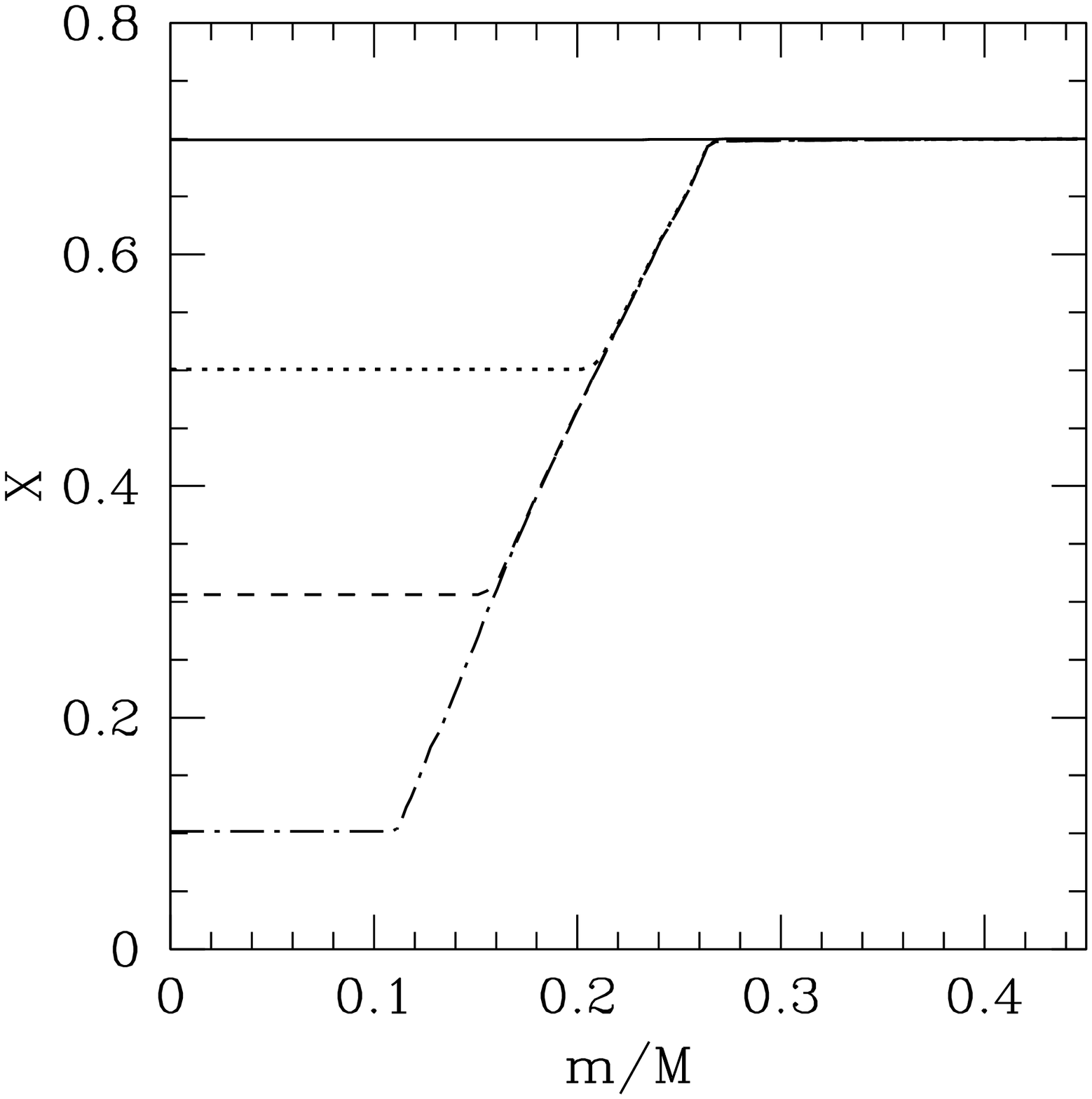}}
\resizebox{0.8\hsize}{!}{\includegraphics[angle=0]{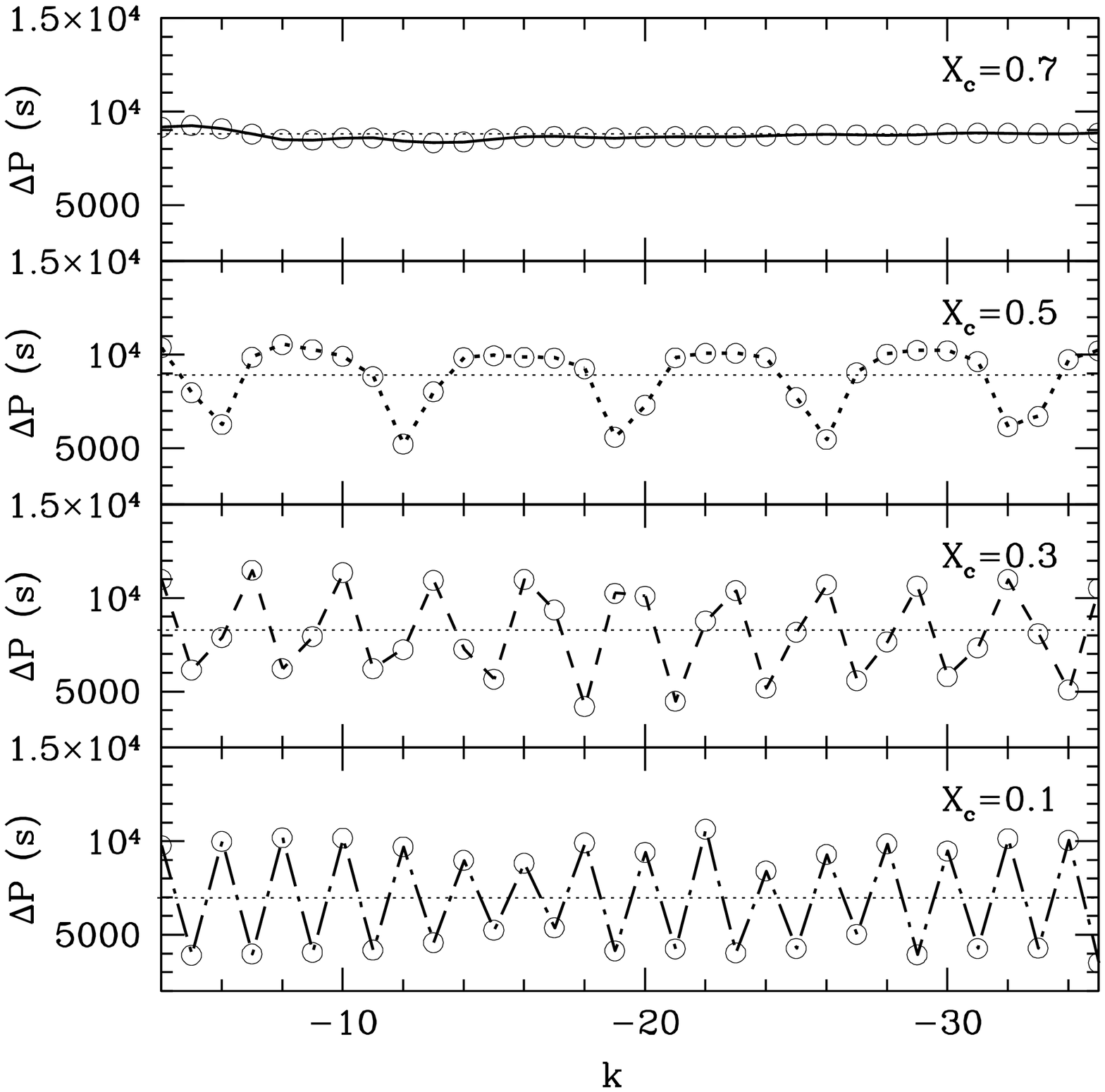}}
\caption{\small Behaviour of the \BV frequency (upper left panel), of the hydrogen abundance profile (upper right panel) and of the $\ell=1$ g-mode period spacing in models of 6 \msun. We consider several models along the main sequence with decreasing central hydrogen abundance ($X_{\rm c}$ 0.7, 0.5, 0.3 and 0.1). In the lower panel the uppermost model has $X_c=0.7$ and the lowermost $X_c= 0.1$.}
\label{fig:6}
 \end{center}
\end{figure}

\begin{figure}
\begin{center}
\resizebox{0.48\hsize}{!}{\includegraphics[angle=0]{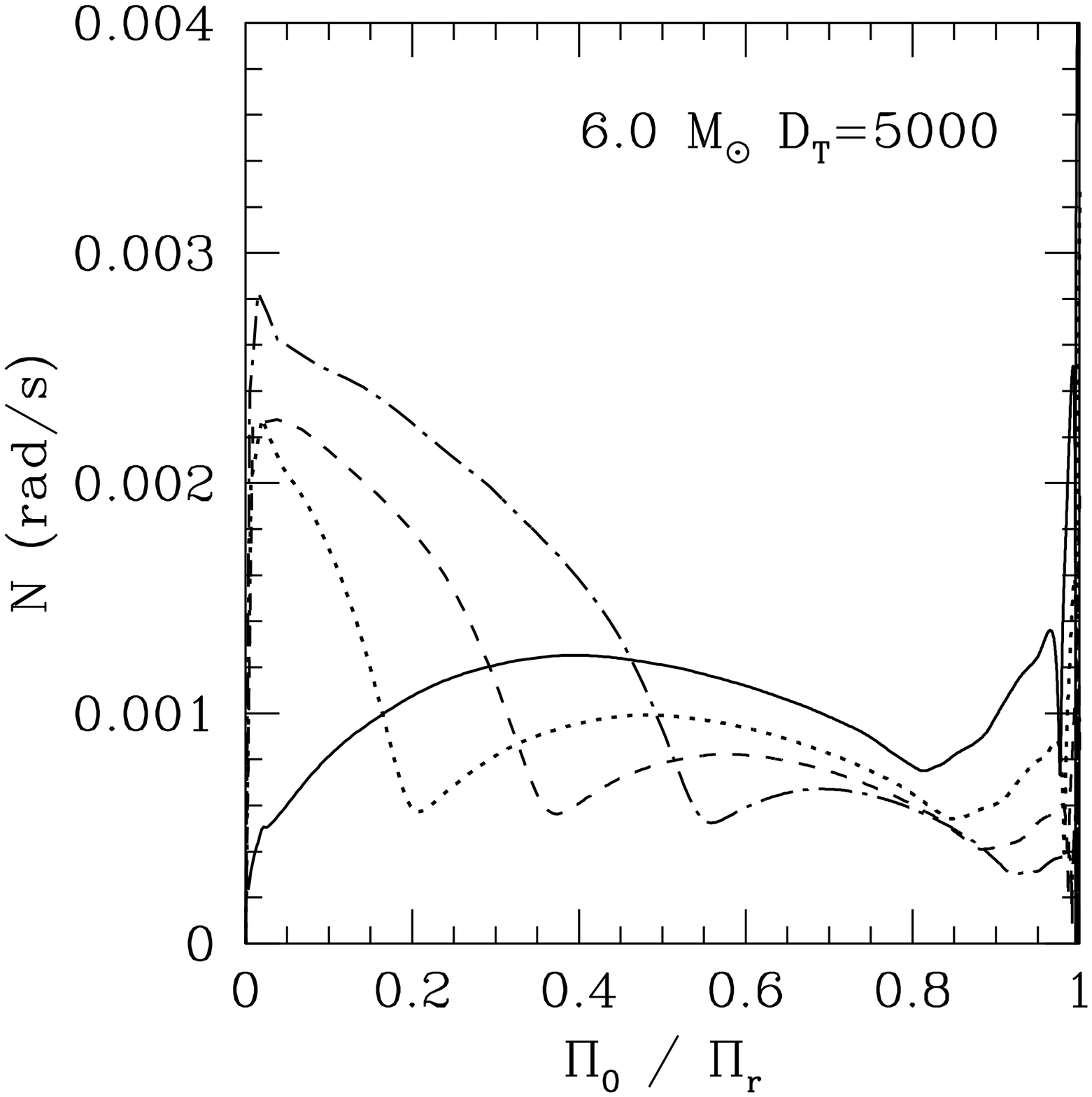}}
\resizebox{0.48\hsize}{!}{\includegraphics[angle=0]{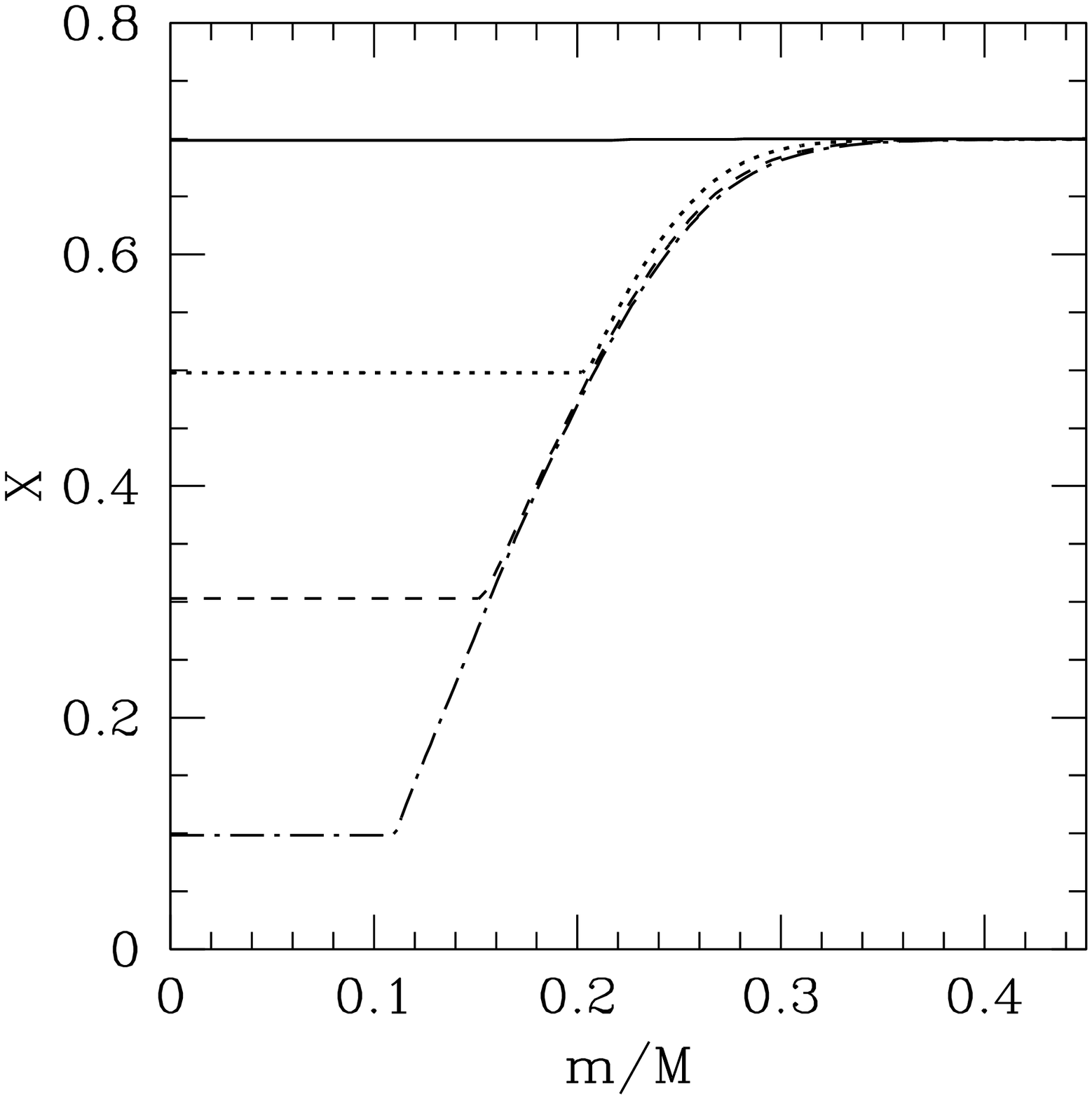}}
\resizebox{0.8\hsize}{!}{\includegraphics[angle=0]{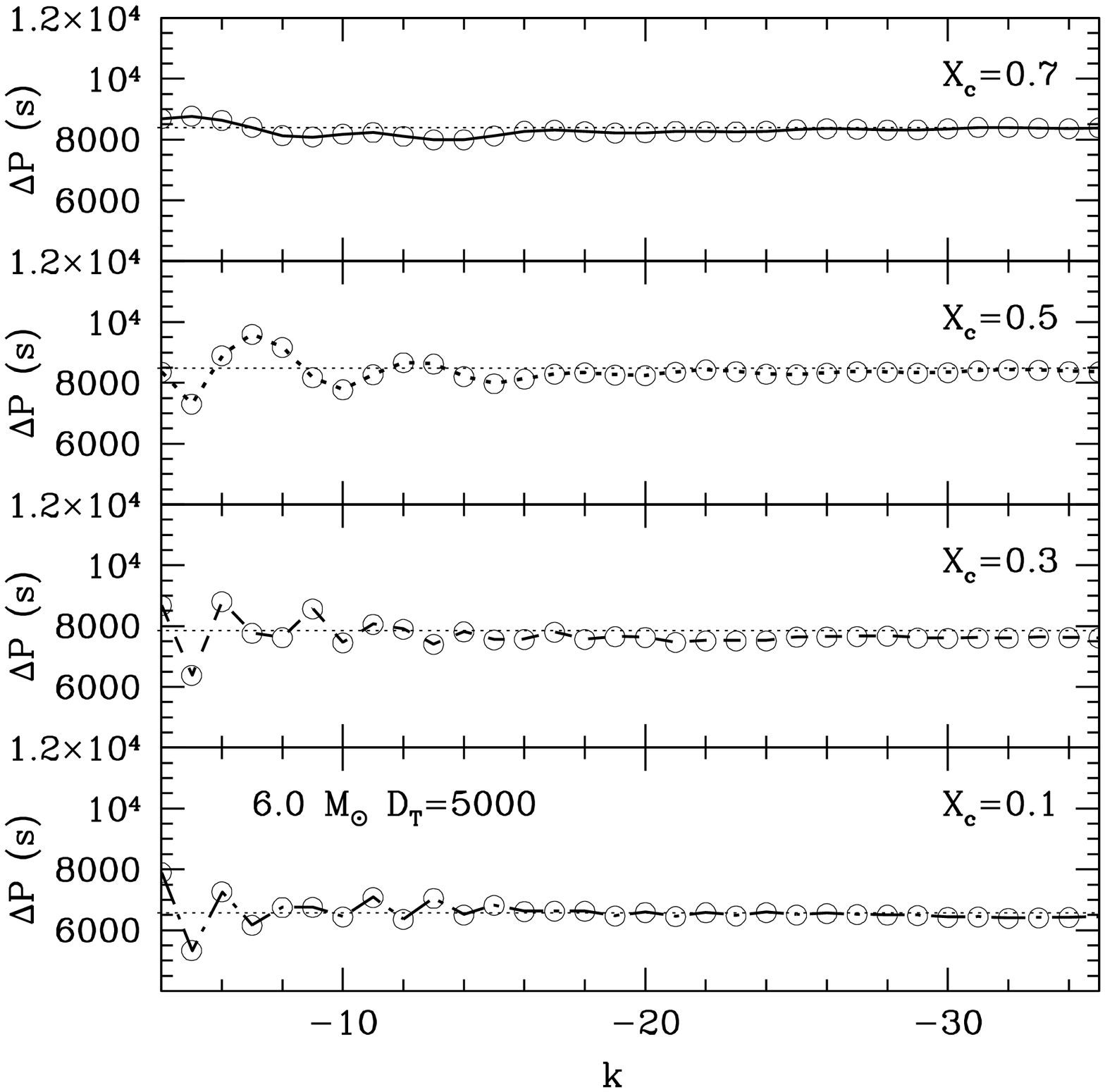}}
\caption{\small Behaviour of the \BV frequency (upper left panel), of the hydrogen abundance profile (upper right panel) and of the $\ell=1$ g-mode period spacing in models of 6.0 \msol computed with a turbulent diffusion coefficient $\rm D_{\rm T}=5000$ cm$^2$ s$^{-1}$.}
\label{fig:6r2}
 \end{center}
\end{figure}

While a detailed description of our investigation is presented in \citet{Miglio07b}, we here recall the results of a simple approximation of g-mode periods that allows to relate the departure from the constant period spacing predicted by the approximation by \citet{Tassoul80}, to the properties of the variation of the \BV frequency located in the $\mu$-gradient region near the core.

This simple approach makes use of the variational principle for adiabatic stellar oscillations \citep[see e.g.][]{Unno89}. The effect of a sharp feature in the model (a chemical composition gradient, for instance) can be estimated
 from the periodic signature in $\delta P$, defined as the difference between the periods of the star showing such a sharp variation and the periods of an otherwise fictitious smooth model.

We consider a model with a radiative envelope and a convective core whose boundary is located at a normalized radius $x_0$. $N_-$ and $N_+$ are the values of the \BV frequency at the outer and inner border of the $\mu$-gradient region. We define $\alpha=\left(\frac{N_+}{N_-}\right)^{1/2}$ with $N_+\leq N_-$. Then $\alpha=1$ describes the smooth model and $\alpha \to 0$, a sharp discontinuity in $N$.

To obtain an estimate of $\delta P$, we adopt \citep[following the approach by][]{Montgomery03} the Cowling approximation, that reduces the differential equations of stellar adiabatic oscillations to a system of the second order. Furthermore, since we deal with high-order gravity modes, the eigenfunctions are well described by their JWKB approximation \citep[see e.g. ][]{Gough93}.

We then model the sharp feature in $\frac{\delta N}{N}$ located at $x=x_\mu$ as:
\begin{equation}
\frac{\delta N}{N}=\frac{1-\alpha^2}{\alpha^2}\, H(x_\mu-x)\;,
\label{eq:step}
\end{equation}
where $H(x)$ is the step function.

Retaining only periodic terms in $\delta P$ and integrating by parts we obtain, after simple calculations \citep[see][]{Miglio07b}:
\begin{equation}
\delta P_k\propto\frac{\Pi_0}{L} \frac{1-\alpha^2}{\alpha^2}\cos{\left(2\,\frac{L\,P_k}{\Pi_\mu}+\phi \right)}\;,
\end{equation}
where $\phi$ is a phase constant; the buoyancy radius of the star is defined as:
\begin{equation}
\label{eq:burad}
\Pi_0^{-1}=\int_{x_0}^1{\frac{|N|}{x'}dx'}\;,
\end{equation}
and the buoyancy radius of the discontinuity is:
\begin{equation}
\Pi_\mu^{-1}=\int_{x_0}^{x_\mu}{\frac{|N|}{x'}dx'}\;\;.
\label{eq:pimu}
\end{equation}

For small $\delta P$ we can substitute in the expression above the asymptotic approximation for g-mode periods derived by \citet{Tassoul80}:
$$P_k=\pi^2\frac{\Pi_0}{L} \left(2k+\phi'\right)\;,$$
where $\phi'$ is a phase constant, and find:
\begin{equation}
\delta P_k\propto\frac{\Pi_0}{L} \frac{1-\alpha^2}{\alpha^2}\cos{\left(2\pi\,\frac{\Pi_0}{\Pi_\mu} k+\phi'' \right)}\;.
\label{eq:varia}
\end{equation}
From this simple approach we derive that the signature of a sharp feature in the \BV frequency is a {\it sinusoidal component in the periods of oscillations}, and thus  in the period spacing, with a periodicity in terms of the radial order $k$ given by
\begin{equation}
\Delta k\simeq\frac{\Pi_\mu}{\Pi_0}\;{.}
\label{eq:variak}
\end{equation}
The amplitude of this sinusoidal component is proportional to the sharpness of the variation in $N$ and does not depend on the order of the mode $k$.

Such a simple approach allows us to easily test the effect of having a less sharp ``glitch'' in the \BV frequency. We model $\delta N$ as a ramp function instead of a step function:
\begin{equation}
\frac{\delta N}{N}=\frac{1-\alpha^2}{\alpha^2}\frac{(x_\mu-x)}{x_\mu-x_0} H(x_\mu-x)\;.
\label{eq:ramp}
\end{equation}

In this case integration by parts leads to a sinusoidal component in $\delta P_k$ whose {\it amplitude is modulated by a factor $1/P_k$} and therefore decreases with increasing $k$, i.e.
\begin{equation}
\delta P_k\propto\frac{1}{P_k}\frac{\Pi_0}{L} \frac{1-\alpha^2}{\alpha^2}\frac{1}{\Pi^{-1}_\mu}\cos{\left(2\pi\,\frac{\Pi_0}{\Pi_\mu} k+\phi' \right)}\;.
\label{eq:senal}
\end{equation}
  The information contained in the amplitude of the sinusoidal component is potentially very interesting. It reflects the different characteristics of the chemical composition gradient resulting, for example, from a different treatment of the mixing process in convective cores and from considering extra-mixing processes acting near the core.

With this simple approximation we can explain the behaviour of the period spacing in main-sequence stars as a  superposition of the constant term derived from Eq. \ref{eq:asy} and periodic components related to $\nabla_\mu$.

As an example we consider in Figure \ref{fig:6} the high-order g-mode spectrum of  6~\msol models along the main sequence.
The amplitude and periodicity of the components in $\Delta P$ can be easily related to the profile of the \BV frequency by means of expression (\ref{eq:variak}).
For instance, the $\Delta k = 7$ in Fig.~\ref{fig:6} (lower panel, $X_{\rm c}=0.50$), corresponds to the sharp signal in $N$ at $\Pi_0/\Pi_{\rm r}\sim 0.14$.
As the star evolves, the sharp feature in $N$ is shifted at higher $\Pi_0/\Pi_{\rm r}$, and when  $\Pi_0/\Pi_\mu \simeq 0.5$ a sort of beating appears in the period spacing due to the fact that the sampling frequency is about half the frequency of the periodic component.

The behaviour of the period spacing is therefore mainly determined by the location and sharpness of the glitch in $N$ in the $\mu$-gradient region. In main-sequence stars the latter, as described in \citet{Miglio07b}, depends on the size of the convective core and on extra-mixing processes (e.g. overshooting, diffusion, turbulent mixing) that may alter the chemical composition profile in the central region of the star.

As an example we consider models of a 6 \msol star computed including as extra mixing turbulent diffusion near the core (described by a turbulent-diffusion coefficient $\rm D_{\rm T}=5\times10^3$ cm$^2$ s$^{-1}$, see Montalb\'an et al., these proceedings for a description of this simple parameterization).
In Fig. \ref{fig:6r2} we notice that such a  mixing has a substantial effect on the period spacing: the amplitude of the periodic components in $\Delta P$ becomes a decreasing function of the radial order $k$. This behaviour can be easily explained by the analytical approximation presented (see Eq.~\ref{eq:senal}), provided the sharp feature in $N$ is modelled not as a step function but, for instance, as a ramp.

\section{Low-order g modes and mixed modes}
\label{sec:low}
Among the different classes of main-sequence pulsating stars, low order gravity modes and mixed-modes are found to be excited in $\beta$ Cephei and $\delta$ Scuti stars. Moreover, in the case of stars evolving as sub-giants, mixed modes are also expected in the frequency domain typical of solar-like oscillations.
\begin{figure}
\begin{center}
\resizebox{.75\hsize}{!}{\includegraphics[angle=0]{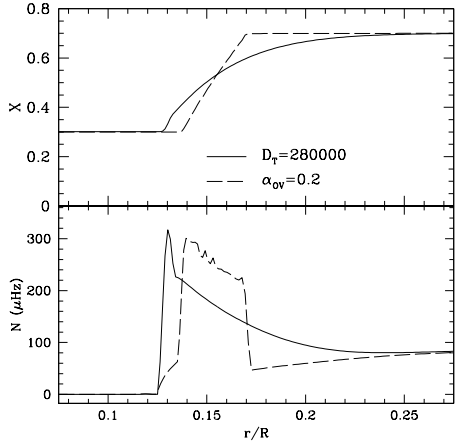}}
\caption{\small Behaviour of the hydrogen abundance profile (upper panel) and of the \BV frequency (lower panel) in 10 \msol models with $X_c \simeq 0.3$. The different lines correspond to models calculated with overshooting ($\alpha_{\rm OV}=0.2$) (dashed) and turbulent mixing with  $\rm D_{\rm T}=2.8\times 10^5$ cm$^2$ s$^{-1}$ (continuous).}\label{fig:s280}
\end{center}
\end{figure}

In the previous section we recalled that the properties of high-order g-mode spectra can be easily related by means of an analytical approximation to the detailed characteristics of the ``glitch'' of $N$ located in the $\mu$ gradient region. This approximation, however, becomes invalid when considering modes of low-order, and cannot be used to give an accurate description of the properties of such modes. A further complication to an analytical description of low-order non-radial modes is given by the interaction between pressure and gravity modes that gives rise to modes of mixed pressure and gravity character (see e.g. \citealt{shibahashi79}).
Nonetheless, as we describe in the following, the analytical description of gravity modes presented in Sec. \ref{sec:high} is still able to qualitatively relate some properties of g-mode spectra for modes of low-order, to the characteristics of the $\mu$-gradient region near the core.

\subsection{$\beta$ Cephei and $\delta$ Scuti stars}
\begin{figure}
\begin{center}
\resizebox{.65\hsize}{!}{\includegraphics[angle=0]{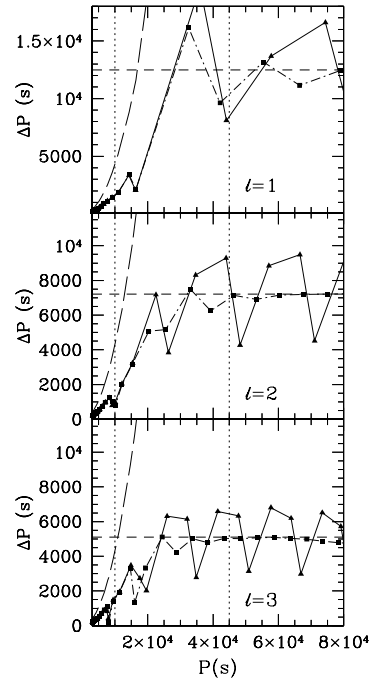}}
\caption{\small Period spacing for $\ell=1,2$ and 3 low order modes (upper, central and lower panel) as a function of the period for 10 \msol models with $ X_c \simeq 0.3$ computed with overshooting $\alpha_{\rm OV}=0.2 $ (triangles connected with solid lines) and with turbulent mixing $\rm D_{\rm T}=2.8\times 10^5$ cm$^2$ s$^{-1}$ (squares connected with dashed-dotted lines). Horizontal short-dashed lines represent the asymptotic value of the period spacing for gravity modes ($\Delta P \propto const$), whereas the long dashed line shows the asymptotic value of $\Delta P$ valid for high order pressure modes ($\Delta P \propto P^2$). The frequency range typical of a $\beta$ Cep star is delimited by vertical dotted lines. }\label{fig:10gl1}
\end{center}
\end{figure}

In main-sequence stars, the frequency of a mixed mode is known to provide information about the extension of the central mixed region.
It has been shown, for instance, that the detection of the frequency of a mixed mode would allow the determination of the overshooting parameter in $\delta$ Scuti stars \citep[see e.g][]{Pamyatnykh99, Michel93, Audard95}. A similar inference is also possible in $\beta$ Cep stars, where the observed oscillation frequencies are now used to place constraints on extra-mixing at the border of the convective core \citep[see e.g.][]{Aerts03, Pamyatnykh04, Mazumdar06, Briquet07}.

Here we investigate, as in the case of high-order g modes, whether the oscillation frequencies typical of $\beta$ Cephei and $\delta$ Scuti stars are sensitive to the detailed shape of the $\mu$ gradient at the edge of the core (and thus to the mixing process that generates such a chemical composition profile).
As in Sec. \ref{sec:high}, we compute models including turbulent mixing near the core, which leads to a smoother $\mu$ gradient (and thus a smoother $N$, see Fig. \ref{fig:s280}) near the boundary of the convective core. We compare the effects of such a mixing with those of overshooting. The models used in the comparison are computed with an overshooting parameter chosen such as to lead to an evolutionary track on the HR diagram close to that obtained with turbulent mixing.

We here compare 10  \msol models with $ X_c \simeq 0.3$ computed with overshooting ($\alpha_{\rm OV}=0.2$) and with turbulent mixing ($\rm D_{\rm T}=2.8\times 10^5$ cm$^2$ s$^{-1}$).
In Fig. \ref{fig:10gl1} we show that in the typical $\beta$ Cep frequency domain  (vertical lines) the difference between the frequencies of these 10 \msol models can still be qualitatively related to the different behaviour of the sharp variation in $N$. 

As shown in the previous section, turbulent mixing leads to a reduction in the amplitude of the periodic components in the period spacing and such a reduction increases with the order of the mode (see Fig. \ref{fig:6r2}).

In Fig. \ref{fig:10gl1} we compare the period spacing for $\ell=1,2$ and 3 modes of a 10 \msol model computed with turbulent mixing or with overshooting. The decrease of the amplitude of the periodic components in $\Delta P$, due to the effect of turbulent mixing, significantly affects the periods of low-order gravity modes and of mixed modes as well. We also compare in Fig. \ref{fig:10gl1} the period spacing  of low-order modes and of mixed modes to the expression of  $\Delta P$ valid in the two asymptotic regimes: for high-order gravity modes ($\Delta P \propto const$) and for high-order pressure modes ($\Delta P \propto P^2$).

We refer to Montalb\'an et al., these proceedings, for a more detailed description of this effect and for a quantitative estimation of the the turbulent mixing induced by rotation in models of typical $\beta$ Cep stars.

\begin{figure}
\begin{center}
\resizebox{.9\hsize}{!}{\includegraphics[angle=0]{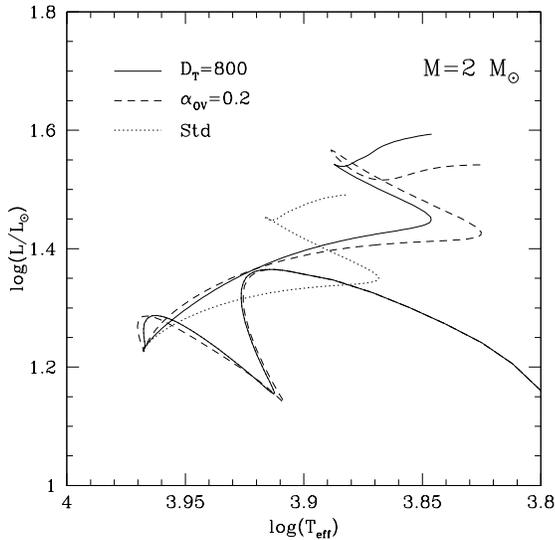}}
\caption{\small  HR diagram showing evolutionary tracks of 2 \msol models calculated without extra-mixing processes (dotted line), with overshooting ($\alpha_{\rm OV}=0.2$, dashed line) and with turbulent mixing ($\rm D_{\rm T}=800$ cm$^2$ s$^{-1}$, full line)}\label{fig:hr2dt}
\end{center}
\end{figure}

\begin{figure}
\begin{center}
\resizebox{.9\hsize}{!}{\includegraphics[angle=0]{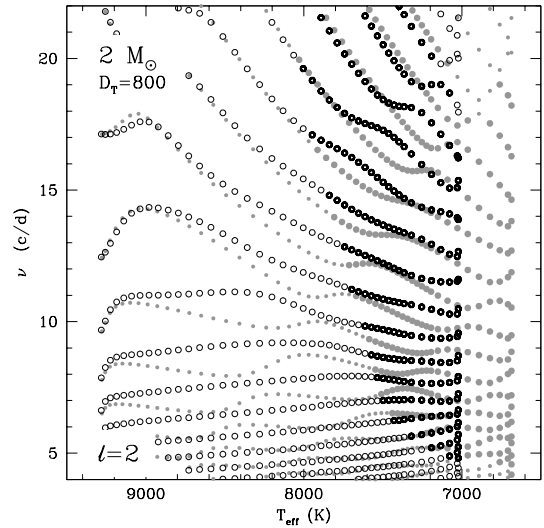}}
\caption{\small  Frequencies of $\ell=2$ pulsation modes as a function of $\,\log{T_{\rm eff}}$ for main-sequence models of a 2 \msol star. Gray dots represent the frequencies of models computed with overshooting ($\alpha_{\rm OV}=0.2$), whereas black circles are the frequencies of models computed with $\rm D_{\rm T}=800$ cm$^2$ s$^{-1}$. Excited modes are represented by larger dots (models with overshooting) and thicker circles (models with turbulent mixing).}\label{fig:2800}
\end{center}
\end{figure}

Though we concentrated our investigations primarily on the effects of turbulent mixing on models of $\beta$ Cephei stars, we here recall that a similar effect is also expected in the case of pulsation modes of $\delta$ Scuti stars. In \citet{Goupil02} it was in fact already suggested that \emph{``[\ldots] extra mixing due to rotation modifies the shape of the inner maximum of N. Signatures of these fine structure in frequencies of mixed and g modes should be detectable.''}.
As an example of this effect  we compare models of 2 \msol stars computed with turbulent mixing ($\rm D_{\rm T}=800$ cm$^2$ s$^{-1}$) and with overshooting ($\alpha_{\rm OV}=0.2$). The corresponding evolutionary tracks in the HR diagram are shown in Fig. \ref{fig:hr2dt}, whereas the behaviour of the oscillation frequencies as a function of $T_{\rm eff}$ is presented in Fig. \ref{fig:2800}. Like in the case of $\beta$ Cep models we notice that turbulent mixing, despite having an effect on the evolutionary tracks similar to that of overshooting, affects differently the properties of pulsation modes.
For a quantitative investigation of this effect, however, comparisons with models where turbulent mixing results, for instance, from a consistent treatment of rotation are still needed.


\subsection{Mixed modes in solar-like oscillations}
\label{sec:solar}
As discussed e.g. in \citet{Dimauro04} and \citet{Miglio07a}, the detection of mixed modes in the spectrum of stars showing solar-like oscillations would allow to determine the evolutionary state of the star and, therefore, to discriminate among theoretical models computed with different values of the overshooting parameter. In fact, in stars that are still in the core-hydrogen  burning phase, the frequencies of gravity modes are not high enough to enter the domain of solar-like oscillations, whereas this is no longer the case in more evolved models, where mixed modes can severely perturb the solar-like oscillation spectrum.

As in the case of more massive stars presented above, we expect that the frequency of a mixed mode depends on the detailed properties of the chemical composition gradient near the core.

\begin{figure}
\begin{center}
\resizebox{.75\hsize}{!}{\includegraphics[angle=0]{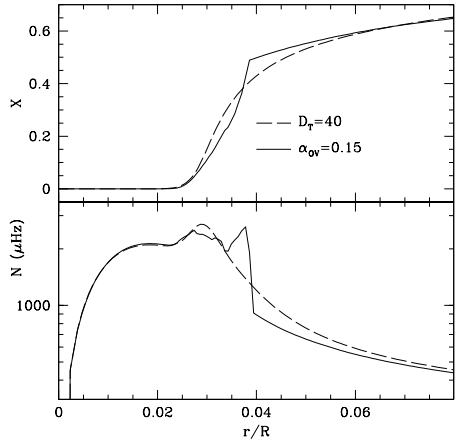}}
\caption{\small  Hydrogen profile (\textit{upper panel}) and \BV frequency (\textit{lower panel}) in models computed with overshooting (continuous line) and with turbulent mixing (long-dashed line)}\label{fig:turbm14}
\end{center}
\end{figure}

\begin{figure}
\begin{center}
\resizebox{.75\hsize}{!}{\includegraphics[angle=0]{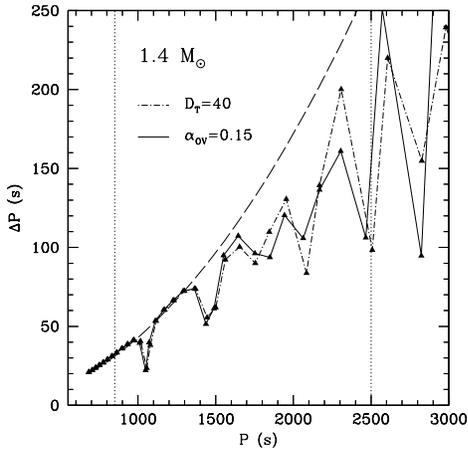}}
\caption{\small Period spacing for $\ell=1$  modes as a function of the period for a 1.4 \msol model computed with overshooting ($\alpha_{\rm OV}=0.15$, solid lines) and with turbulent mixing ($\rm D_{\rm T}=40$ cm$^2$ s$^{-1}$, dashed-dotted lines). The long-dashed line represents the asymptotic value of $\Delta P$ valid for high order pressure modes. The frequency range of solar-like oscillations is delimited by vertical dotted lines. }\label{fig:turbm14f}
\end{center}
\end{figure}

We here consider, as an example,  two models of 1.4 \msol subgiant stars with the same surface properties ($L$,$T_{\rm eff}$) but that are computed including different extra mixing processes in the core: overshooting ($\alpha_{\rm OV}$=0.15) and turbulent mixing ($\rm D_{\rm T}=40$ cm$^2$ s$^{-1}$). Even if the models have already left the main sequence, we see in Fig.\ref{fig:turbm14} that their chemical composition profile (upper panel) still bears the signature of a different mixing process in the core, which also affects the \BV frequency (lower panel). In analogy to Fig. \ref{fig:10gl1}, we show in Fig. \ref{fig:turbm14f} that the differences in the frequencies of gravity modes, due to a different behaviour of $N$, can also affect the spectrum in the domain of solar-like oscillations (vertical dotted lines). A more detailed discussion on this effect and, in particular, on its magnitude and detectability for the specific case of the binary 12 B\"ootis is given in \citet{Miglio07a}.


\section{Conclusions}
We have shown that the periods of gravity modes in main-sequence stars can be related, by means of analytical approximations,  to the detailed characteristics of the $\mu$-gradient region that develops near the energy generating core and thus to the mixing processes that affect the behaviour of $\mu$ in the central regions.

Further investigations are however needed to assess under which observational conditions such information can be recovered from the oscillation frequencies, given realistic observational errors and, in the case of SPB and $\gamma$ Dor pulsators, given the severe influence of rotation on the spectrum of g modes.
\acknowledgements
A.M. and J.M. acknowledge financial support from the European Helio- and Asteroseismology
Network HELAS, from the Prodex-ESA Contract Prodex 8 COROT (C90199) and from FNRS. P.E. is thankful to the Swiss
National Science Foundation for support.
\bibliographystyle{aa}
\bibliography{miglio}





\end{document}